%% file: main.tex
\documentclass[runningheads]{llncs}
\usepackage[T1]{fontenc}
\usepackage{float}
\usepackage{caption}
\usepackage{graphicx} % For resizing the 
\usepackage{tabularx}
\usepackage{booktabs} % For \toprule, \midrule, and \bottomrule
\usepackage{subcaption}
\usepackage{hyperref} 

\begin{document}
\titlerunning{AutoPET III challenge}
\title{Autopet III challenge: Incorporating anatomical knowledge into nnUNet for lesion segmentation in PET/CT}
\author{Hamza Kalisch\inst{1,2} \and
Fabian Hörst\inst{1,3,5} \and
Ken Herrmann\inst{2} \and
Jens Kleesiek\inst{1,3,4,5} \and
Constantin Seibold\inst{1,2}}
\authorrunning{H. Kalisch et al.}
\institute{Institute for AI in Medicine (IKIM), University Hospital Essen (AöR), Essen, Germany \and
Department of Nuclear Medicine, University Hospital Essen (AöR), Essen, Germany
\and 
Cancer Research Center Cologne Essen (CCCE), West German Cancer Center Essen University Hospital Essen (AöR), Essen, Germany
\and
German Cancer Consortium (DKTK, Partner site Essen), Heidelberg, Germany
\and
Department of Physics, TU Dortmund, Dortmund, Germany\\
\email{\{firstname.lastname\}@uk-essen.de}
}

\maketitle              % typeset the header of the contribution

\begin{abstract}
Lesion segmentation in PET/CT imaging is essential for precise tumor characterization, which supports personalized treatment planning and enhances diagnostic precision in oncology. However, accurate manual segmentation of lesions is time-consuming and prone to inter-observer variability. 
Given the rising demand and clinical use of PET/CT, automated segmentation methods, particularly deep-learning-based approaches, have become increasingly more relevant. The autoPET III Challenge focuses on advancing automated segmentation of tumor lesions in PET/CT images in a multitracer multicenter setting, addressing the clinical need for quantitative, robust, and generalizable solutions. 
Building on previous challenges, the third iteration of the autoPET challenge introduces a more diverse dataset featuring two different tracers (FDG and PSMA) from two clinical centers.
To this extent, we developed a classifier that identifies the tracer of the given PET/CT based on the Maximum Intensity Projection of the PET scan. We trained two individual nnUNet-ensembles for each tracer where anatomical labels are included as a multi-label task to enhance the model's performance. Our final submission achieves cross-validation Dice scores of 76.90\% and 61.33\% for the publicly available FDG and PSMA datasets, respectively. The code is available at \href{https://github.com/hakal104/autoPETIII/}{https://github.com/hakal104/autoPETIII/}
\end{abstract}

\section{Introduction}

\input{Introduction}

\section{Method}

\input{Method}

\section{Experimental setup and evaluation}

\input{setup}

\section{Results}

\input{Results}

\section{Conclusion}

In this work, we addressed the task of automated lesion segmentation in PET/CT imaging within the context of the autoPET III challenge. Our approach involved developing a classifier to identify the tracer type based on the PET scan's Maximum Intensity Projection (MIP). We then trained separate nnU-Net ensembles for each tracer, integrating anatomical labels as a weighted multi-label classification task to enhance segmentation performance.
Our methodology tackled the inherent difficulties of distinguishing between physiological and pathological uptake patterns, given the distinct tracer-specific uptake characteristics and varied imaging protocols.
Our final models achieved Dice scores of 76.90\% for the FDG dataset and 61.33\% for the PSMA dataset, representing a significant improvement over a baseline nnU-Net model trained on both datasets. Additionally, our approach maintained lower volumes of false negatives and false positives, underscoring the effectiveness of incorporating tracer-specific classification and anatomical knowledge into the segmentation process. A drawback of employing two separate models in the context of the challenge is that the effectiveness of this approach depends heavily on the tracer classifier's accuracy. However, in a clinical setting, the tracer type is known at inference time, so this issue is mitigated.

\section*{Acknowledgements}
This work received funding from \lq KITE' (Plattform für KI-Translation Essen) from the REACT-EU initiative (\url{https://kite.ikim.nrw/}, EFRE-0801977) and the Cancer Research Center Cologne Essen (CCCE).

\bibliographystyle{plain}
\bibliography{mybib}
\end{document}

%% file: Introduction.tex
Positron Emission Tomography (PET) combined with Computed Tomography (CT) imaging is an essential tool in oncology, providing detailed information about tumors and helping to personalize treatment plans \cite{Farwell2014,Schwenck2023}. While the CT scan offers detailed anatomical information, PET imaging complements it by highlighting the metabolic activity of different cancers through the use of different tracers such as 18F-Fluorodeoxyglucose (FDG) and Prostate-Specific Membrane Antigen (PSMA). FDG is valued for its ability to detect metabolic activity in various cancers, while PSMA is particularly useful for targeting prostate cancer due to its specific binding to prostate tumor cells \cite{jochumsen2024psma,Almuhaideb2011}. 

Lesion segmentation in PET/CT imaging is essential for detailed tumor analysis and boosting diagnostic precision. However, manual lesion segmentation in PET/CT scans is still time-consuming and prone to inconsistencies, making automated methods increasingly important for improving diagnostic precision and efficiency in clinical settings.
The task is however subject to differences in patient physiology, tracer-specific uptake patterns, and diverse imaging protocols, making it challenging to distinguish between physiological and pathological uptake. 

The autoPET III challenge addresses the need for robust and generalizable segmentation solutions by focusing on automated lesion segmentation in PET/CT scans across multiple tracers and clinical settings. It builds on past efforts by adding a new PSMA dataset \cite{jeblick2024psmapetct} of 597 whole-body PET/CT studies alongside the FDG dataset \cite{gatidis2022fdgpetct} of 1,014 studies, which was used in the previous two autoPET challenges. The FDG dataset was acquired at the University Hospital Tübingen (UKT), while the PSMA dataset was obtained from the LMU Hospital in Munich. 

Previous challenge results \cite{isensee2021nnu} have extensively shown the effectiveness of the nnU-Net \cite{gatidis2023autopet} for the task of lesion segmentation in FDG PET/CTs. However, the distinct uptake distributions of FDG and PSMA tracers present additional challenges, complicating the differentiation between physiological and pathological uptake. To address this, we trained a classifier to distinguish between FDG and PSMA tracers and enhanced the model’s accuracy by integrating anatomical knowledge through a multilabel approach \cite{murugesan2023evaluating,murugesan2023improving,jaus2024anatomy}. This involves training two separate nnUNet ensembles on the FDG and PSMA datasets, where anatomical masks are included alongside lesion labels in a multi-class classification task. 

%% file: Method.tex
\subsection{Pipeline overview}

\begin{figure}[ht]
    \centering
    \includegraphics[width=1\textwidth]{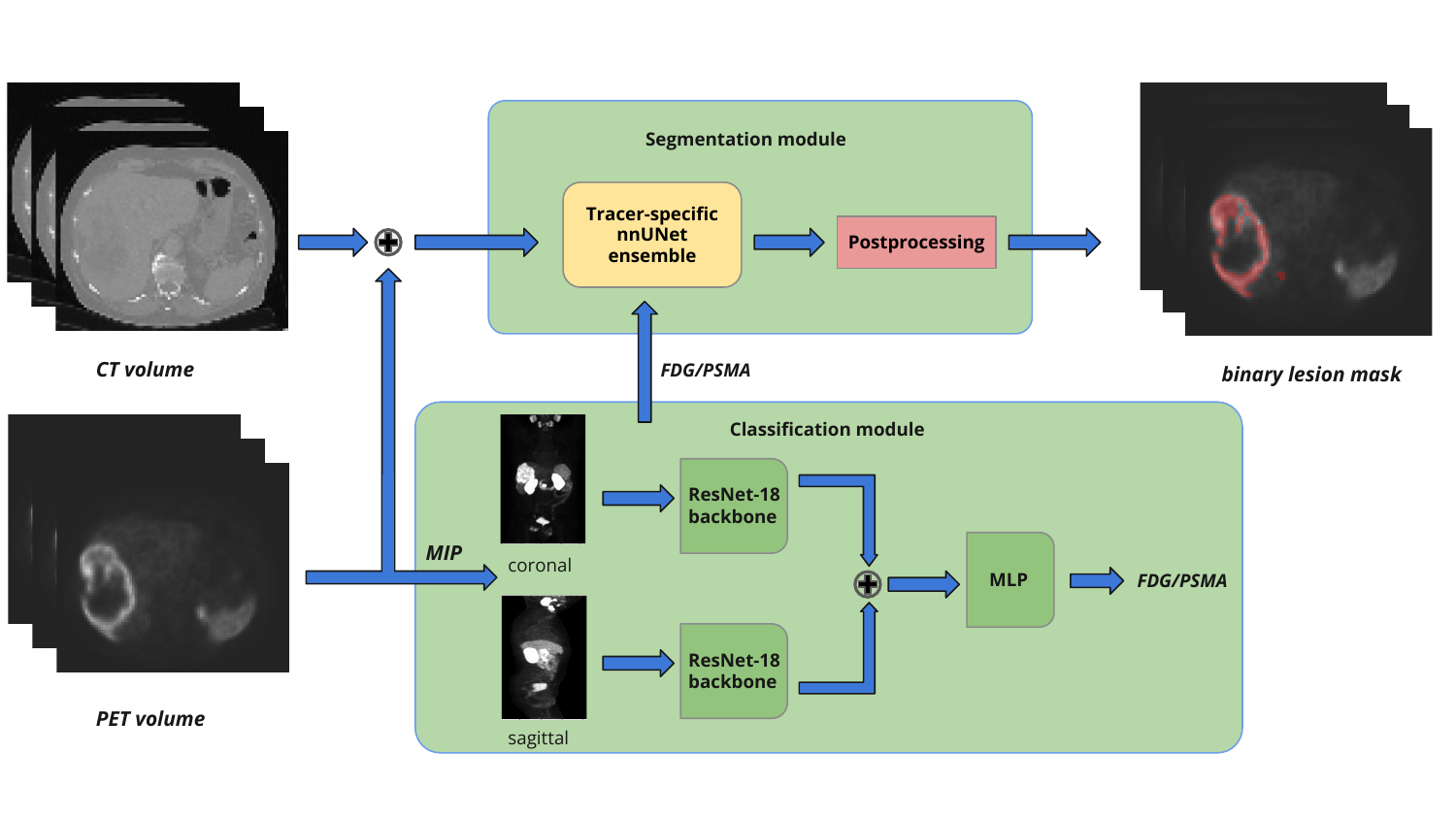}  % Replace with your image file
    \caption{Pipeline of our final submission. In the first step, the PET volume from a given PET/CT scan is input into a classification module that determines the tracer type. PET and CT volumes are then concatenated and fed into a nnU-Net ensemble that was trained on the dataset of the classified tracer. The final binary lesion mask is output after postprocessing.}
    \label{fig:method}
\end{figure}

Fig. \ref{fig:method} illustrates the workflow of our final submission. Given a PET/CT scan, the PET volume is first input into a classification module that identifies the tracer. In the next step, both the PET and the CT are concatenated and processed through a tracer-specific nnUNet-ensemble, which generates the initial segmentation mask. After applying post-processing steps, the final segmentation mask is produced. In the following, the classification module is described in detail.

\subsection{MIP-based tracer classification}
\label{sec:classifier}
\begin{figure}[ht]
    \centering
    \begin{subfigure}{0.45\textwidth}
        \centering
        \includegraphics[width=\linewidth]{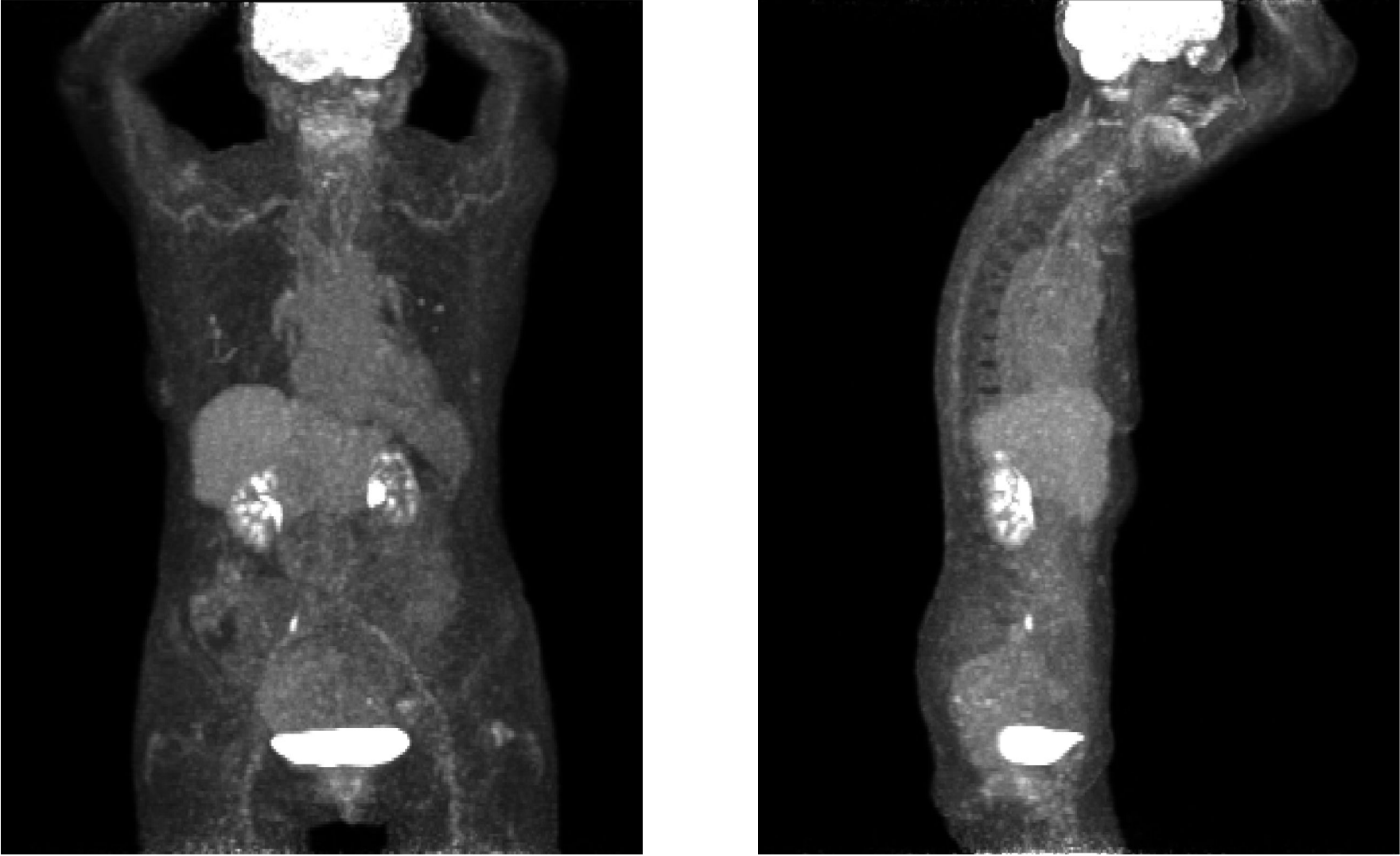}
        \caption{FDG}
        \label{fig:figure1}
    \end{subfigure}
    \hfill
    \begin{subfigure}{0.45\textwidth}
        \centering
        \includegraphics[width=\linewidth]{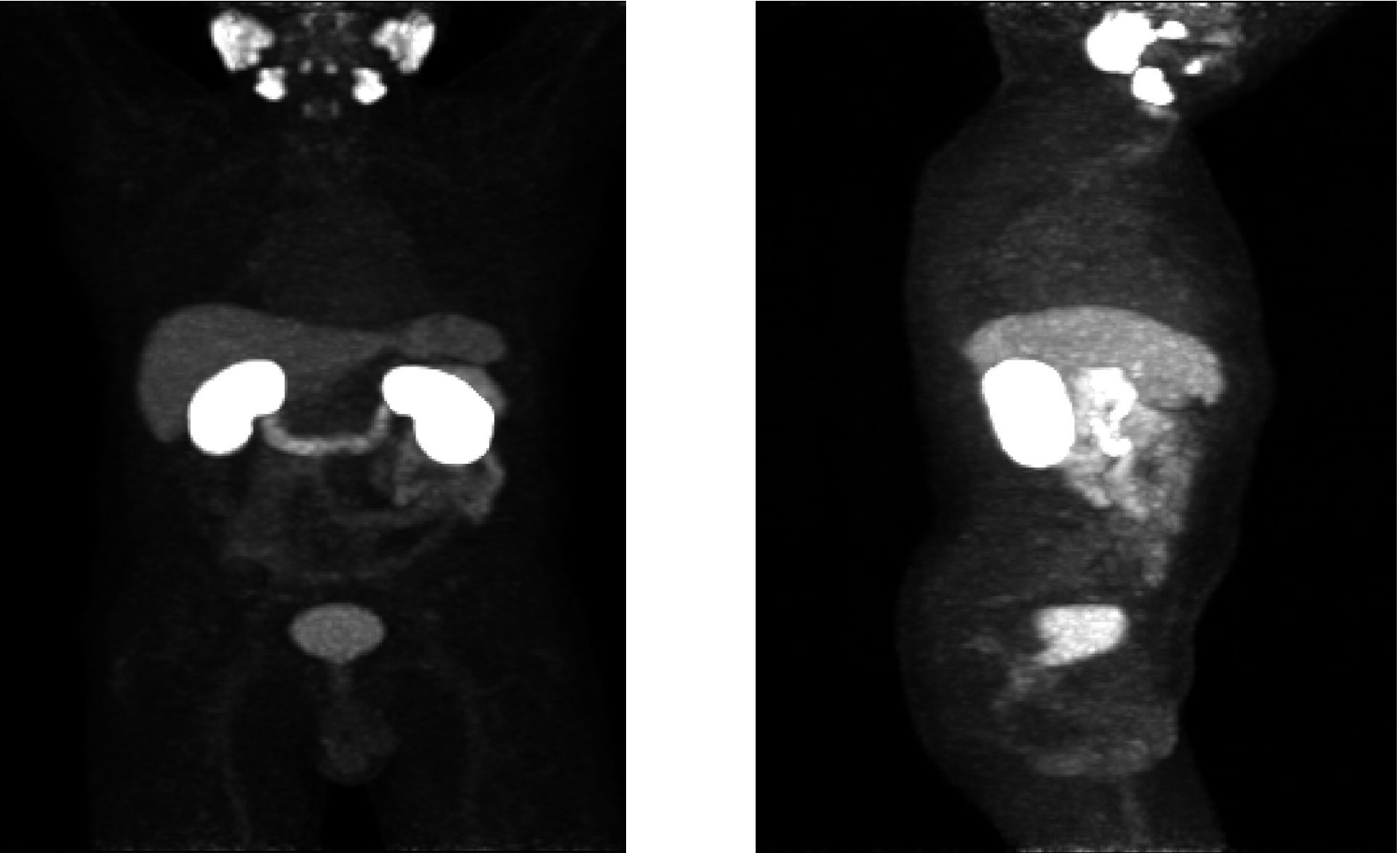}
        \caption{PSMA}
        \label{fig:figure2}
    \end{subfigure}
    \caption{Exemplary coronal and sagittal MIPs for PET scans of two healthy patients using the FDG tracer and PSMA tracer. Whiter regions represent higher SUV values. In both views, the FDG MIP highlights clear higher uptake patterns for the brain and urinary bladder, whereas the latter reveals higher uptake in the kidneys, submandibular glands, and parotid glands.}
    \label{fig:mips}
\end{figure}

%The goal of the classification module is to solve the binary %classification task
Due to the distinct mechanisms of FDG and PSMA  as PET tracers, their uptake distributions exhibit notable differences. The most notable difference is observed in the brain, where FDG demonstrates significantly higher SUV values than PSMA. Additionally, FDG PET scans generally show lower average uptake in the kidneys and liver, whereas PSMA demonstrates significant uptake in the submandibular and parotid glands. To capture these uptake distributions while simplifying image complexity, we utilize Maximum Intensity Projections (MIP) of the PET scans in the coronal and sagittal plane as input to our classification model. This approach significantly reduces inference time by minimizing the complexity of the input data compared to using the full 3D volume. Fig. \ref{fig:mips} illustrates the distinct uptake patterns between healthy FDG and PSMA scans as observed in the MIPs. The differences in uptake patterns between tracers also vary with pathological conditions. Since PSMA scans are used in the context of prostate cancer, its bone metastases are more commonly observed in PSMA scans, while they may be less visible in FDG scans.

We first train two separate ResNet18 models on coronal and sagittal MIPs. Next, a multilayer perceptron (MLP) is trained on top of the concatenated features from the frozen backbones of both models. The MLP outputs a binary classification indicating whether the used tracer in the given PET/CT scan is FDG or PSMA.

%- images of healthy FDG / PSMA MIP patients, to show different distributions  (coronal and sagittal)
%- example images of ill patients (MIPS) FDG vs PSMA, different lesion distribution

\subsection{Anatomy-guided lesion segmentation}
\label{sec:ana_method}
For the segmentation of the lesions, we use the nnU-Net framework. To better distinguish between physiological and pathological uptake, we guide the network to learn organ-specific uptake patterns by incorporating anatomical knowledge.
To this extent, we explore how anatomical structures can further improve the initial results. 117 different anatomical labels are extracted using TotalSegmentator \cite{wasserthal2023totalsegmentator} for each CT volume of the datasets. Besides using all anatomical structures, we also take different subsets depending on the dataset which are described in Sec. \ref{sec:preprocessing}. We train and evaluate models for each dataset using the following three approaches:

\begin{itemize}
    \item \textbf{Multilabel classification}: Anatomical structures are used as additional label input alongside the lesion labels. Additionally, we introduce a weighting factor $\lambda$ to adjust the relative importance of lesion labels compared to anatomical labels within the Dice Cross-Entropy loss function.
    
    \item \textbf{Channel concatenation}: Anatomical labels are added as an additional channel to the PET/CT data, resulting in a three-channel input for the model.
    
    \item \textbf{Pretraining}: We finetune the STU-Net \cite{huang2023stu} architecture, which has been pretrained on the TotalSegmentator dataset, on only lesion labels as well as the mentioned subsets of additional anatomical labels.
\end{itemize}

%% file: setup.tex
\subsection{Datasets}

Experiments were conducted on two datasets, which we will simply refer to as the FDG dataset and the PSMA dataset. The FDG dataset consists of 1,014 whole-body PET/CT studies from 900 patients, acquired at the University Hospital Tübingen using a Siemens Biograph mCT scanner. It includes 501 studies from patients diagnosed with malignant melanoma, lymphoma, or lung cancer, along with 513 studies from negative control patients.
The PSMA dataset includes 597 whole-body PET/CT studies of male patients with prostate carcinoma, comprising 537 studies with PSMA-avid tumor lesions and 60 studies without. The studies were acquired from the LMU Hospital in Munich using three different PET/CT scanners (Siemens Biograph 64-4R TruePoint, Siemens Biograph mCT Flow 20, and GE Discovery 690). It should be noted that for the segmentation experiments in Sec. \ref{sec:baseline} and Sec. \ref{sec:ana} redacted samples were utilized.

\subsection{Evaluation protocol and metrics}

\paragraph{\textbf{Tracer classification.}} We merged the FDG and PSMA datasets and performed an 80/20 train-test split. To evaluate model robustness, we conducted 5-fold cross-validation on the training set. Classification accuracy is reported across both the cross-validation folds and the independent test set.

\paragraph{\textbf{Lesion segmentation.}} For each dataset, five samples were excluded and used as an additional internal test set. We then conducted 5-fold cross-validation using nnUNet, following the training strategies outlined in Sec. \ref{sec:ana_method} and report results based on three different metrics. In addition to the Dice score, the evaluation metrics of the autoPET challenge include false positive and false negative lesion volumes. The false positive volume (FPV) refers to the volume of predicted connected components that do not overlap with any ground truth components, while the false negative volume (FNV) measures the ground truth connected components that are missed by the predicted segmentation mask.

\subsection{Preprocessing and implementation details}
\label{sec:preprocessing}

\paragraph{\textbf{Tracer classification.}} The coronal and sagittal MIPs were resized to a size of 224 x 224 and given to the ResNet18 as inputs. During training, we applied various data augmentations such as horizontal flipping, random rotation, Gaussian blurring, and cropping. The ResNets are trained for 50 epochs with a learning rate and weight decay of 0.0005 using ADAM as an optimizer, while the MLP is trained for 20 epochs with a learning rate of 0.0001.

\paragraph{\textbf{Lesion segmentation.}} For both the PET and the CT, global dataset percentile clipping as well as normalization based on the mean and standard deviation of the dataset were performed. Notably, for the final submission, we applied z-score normalization specifically to the PET scans.
We trained standard 3D full-resolution U-Net models, using the default configurations and data augmentations from nnUNet. This includes a range of augmentations such as Gaussian noise, random rotation, cropping, Gaussian blur, down-sampling, and gamma correction. We trained for 1000 epochs with a batch size of 2 and an initial learning rate of 0.01. For fine-tuning the STU-Net model and concatenating the labels as a third channel, we adjusted the learning rate to 0.001.
For approaches that utilize anatomical labels, we mainly considered organs from healthy patients that show a mean SUV uptake of higher than 2. For FDG, these labels are given by brain, urinary bladder, spleen, liver, heart, kidneys, duodenum, prostate, small bowel, and esophagus. Besides the colon and pancreas, we also added the lungs to this subset, since including them has shown a performance boost. For the PSMA models, we included the gallbladder and excluded the brain, lungs, and colon.

%% file: Results.tex
\subsection{Baseline models}
\label{sec:baseline}

Before presenting the results for tracer classification and evaluating different lesion segmentation algorithms, we first compare the performance of a simple baseline nnU-Net model trained on a combined FDG and PSMA dataset with models trained individually on each dataset. As shown in Tab. \ref{tab:baseline_results}, the results indicate that the Dice score and FNV for the FDG dataset are slightly better with the separate training approach, while the FPV is worse. In contrast, the model trained individually on the PSMA dataset significantly outperforms the combined model in both Dice score and FPV, with similar performance in FNV.
These findings highlight the benefits of training separate models for each dataset, thereby justifying the development of a dedicated tracer classification module.

\begin{table*}[t]
\centering
\small % Ensure the same font size here
\begin{tabular}{llrrrrrrr}
\toprule
\textbf{Method} & & \multicolumn{3}{c}{\textbf{FDG Dataset}} & & \multicolumn{3}{c}{\textbf{PSMA Dataset}} \\
\cmidrule(lr){3-5} \cmidrule(lr){6-8}
 & & \textbf{DICE} & \textbf{FPV} & \textbf{FNV} & &\textbf{DICE} & \textbf{FPV} & \textbf{FNV} \\
\midrule
\textbf{Combined training} & & 0.6483 & \textbf{18.93} & 12.18 & &0.5258 & 16.33 & \textbf{19.46} \\
\textbf{Individual training} & & \textbf{0.6538} & 25.34 & \textbf{9.72} & & \textbf{0.6014} & \textbf{12.17} & 20.49 \\
\bottomrule
\end{tabular}%
\vspace{0.5cm}
\caption{Averaged 5-fold cross-validation results for the combined and individual training approach on the FDG and PSMA dataset.}
\label{tab:baseline_results}
\end{table*}

\subsection{Tracer classification}

Tab. \ref{tab:transposed_evaluation_metrics} shows the final results for tracer classification using the fusion approach described in Sec. \ref{sec:classifier} . We compare this approach with two models trained separately on coronal and sagittal MIPs. All methods demonstrate high effectiveness, achieving accuracies of at least 99\%. This high performance suggests that the classification task is relatively simple due to the clear distinction between tracer distributions. Although all methods attain perfect accuracy on the test set, the fusion approach outperforms the individual MIP-based models for cross-validation. This approach benefits from integrating complementary information from both MIPs, leading to more reliable and accurate classification results. Given that the additional computational cost of including both MIPs is minimal compared to the lesion segmentation model, its use is well-justified.

\begin{table*}[t]
\centering
\small % Ensure the same font size here
\begin{tabular}{lccc|ccc}
\toprule
\textbf{Method} & \textbf{5-Fold CV} & \textbf{Test Dataset} \\
\midrule
\textbf{Coronal} & 0.9977 & 1.0000 \\
\textbf{Sagittal} & 0.9977 & 1.0000  \\
\textbf{Fusion} & 0.9992 & 1.0000  \\
\bottomrule
\end{tabular}%
\vspace{0.5cm}
\caption{Classification accuracies for the 5-Fold Cross-Validation and test dataset using ResNets trained separately on coronal MIPs, sagittal MIPs, and the fusion approach. }
\label{tab:transposed_evaluation_metrics}
\end{table*}

\subsection{Anatomy-guided lesion segmentation}
\label{sec:ana}

The cross-validation results on the FDG and PSMA datasets using the various lesion segmentation approaches described in Sec. \ref{sec:ana_method} are shown in Tab. \ref{tab:ana_results}. The baseline models refer to the nnU-Net models that are trained individually on the FDG and PSMA datasets. Compared to the baseline every approach shows improvements for every metric except the multilabel classification model that is trained on all anatomical labels. We suggest that the Dice is even slightly lower than the baseline because the model places less emphasis on the lesion class during training. By taking a relevant subset of organs and therefore also fewer organs the Dice score improves dramatically. Using the weighted approach further increases the Dice score.
The weighted multilabel approach is also the best-performing method on the PSMA dataset while keeping FPV and FNV comparable to other methods.

For the finetuning of the pretrained STU-Net model using the multilabel approach, we surprisingly find lower performance in the Dice score. It should be noted that this might be attributed to the use of a lower learning rate. Specifically, when training on the PSMA dataset using the multilabel approach with a learning rate of 0.001, we also observe a similar decrease in performance.

\begin{table*}[t]
\centering
\small % Ensure the font size is applied here
\begin{tabular}{lrrrrrrr}
\toprule
\textbf{Method} & & \multicolumn{3}{c}{\textbf{FDG Dataset}} & \multicolumn{3}{c}{\textbf{PSMA Dataset}} \\
\cmidrule(lr){3-5} \cmidrule(lr){6-8}
 & & \textbf{DICE} & \textbf{FPV} & \textbf{FNV} & \textbf{DICE} & \textbf{FPV} & \textbf{FNV} \\
\midrule
\textbf{Baseline} & & 0.6583 & 25.34 & 9.72 & 0.6014 & 12.17 & 20.49 \\
\midrule
\textbf{Multilabel Classification} & & & & & & & \\
\quad All labels & & 0.6530 & 9.64 & 7.96 & 0.5343 & \textbf{9.82} & \textbf{10.55} \\
\quad Subset of Labels & & 0.7346 & 7.36 & 6.39 & 0.5968 & 10.64 & 12.68 \\
\quad Weighted Approach ($\lambda=3$) & & \textbf{0.7493} & 7.69 & 5.90 & \textbf{0.6133} & 12.97 & 11.23 \\
\midrule
\textbf{Channel Concatenation} & & & & & & & \\
\quad All labels & & 0.7193 & 9.19 & 7.79 & 0.5930 & 11.04 & 12.79 \\
\midrule
\textbf{Pretraining} & & & & & & & \\
\quad Single label & & 0.7442 & 8.52 & \textbf{5.66} & 0.5909 & 11.73 & 17.44 \\
\quad Multilabel & & 0.7244 & \textbf{6.50} & 7.40 & 0.5784 & 10.95 & 12.54 \\
\bottomrule
\end{tabular}%

\vspace{0.5cm}
\caption{Cross-validation results for various lesion segmentation approaches using anatomical information on the FDG and PSMA dataset.}
\label{tab:ana_results}
\end{table*}

\subsection{Postprocessing}

For postprocessing, we evaluate the impact of thresholding methods on the segmentation performance for two scenarios.

Tab. \ref{tab:cc} examines the effect of removing connected components smaller than a specified threshold on the performance metrics for both datasets. The predicted segmentation masks from the cross-validation results of the final submission models (see Sec. 4.5) are used for evaluation. Based on the result that this method even leads to significant increases in FNV for a threshold of 1, we don't use this method in our final submission. 

Additionally, we assess how varying SUV thresholds impact the model's performance (refer to Tab. \ref{tab:suv}). This involves setting the predicted segmentation masks to 0 for PET values that fall below a designated threshold. According to our findings, we adopt SUV thresholds of 1.5 for the FDG model and 1 for the PSMA model.
\begin{table*}[t]
\centering
\small % Ensure the same font size here
\begin{tabular}{lrrrrrr}
\toprule
\textbf{CC threshold} & \multicolumn{3}{c}{\textbf{FDG Dataset}} & \multicolumn{3}{c}{\textbf{PSMA Dataset}} \\
\cmidrule(lr){2-4} \cmidrule(lr){5-7}
 & \textbf{DICE} & \textbf{FPV} & \textbf{FNV} & \textbf{DICE} & \textbf{FPV} & \textbf{FNV} \\
\midrule
\textbf{1} & 0.0000 & 0.0000 & 0.1059 & 0.0000 & -0.0271 & 0.2028 \\
\textbf{2} & 0.0000 & -0.0123 & 0.1708 & -0.0005 & -0.0629 & 0.4956 \\
\textbf{3} & 0.0000  & -0.0196 & 0.1994 & -0.0007 & -0.1024 & 0.8054 \\
\textbf{5} & 0.0001  & -0.0354 & 0.2467 & -0.0019 & -0.1852 & 1.2753 \\
\textbf{10} & -0.0007 & -0.0739 & 0.3834 & -0.0074 & -0.4350 & 3.1680 \\
\bottomrule
\end{tabular}%
\vspace{0.5cm}
\caption{Absolute differences for the validation results of the FDG and PSMA models when all connected components in the predicted segmentation mask with a length smaller than a specified threshold are set to zero.}
\label{tab:cc}
\end{table*}

\begin{table*}[t]
\centering
\small % Ensure the same font size here
\begin{tabular}{lrrrrrr}
\toprule
\textbf{SUV threshold} & \multicolumn{3}{c}{\textbf{FDG Dataset}} & \multicolumn{3}{c}{\textbf{PSMA Dataset}} \\
\cmidrule(lr){2-4} \cmidrule(lr){5-7}
 & \textbf{DICE} & \textbf{FPV} & \textbf{FNV} & \textbf{DICE} & \textbf{FPV} & \textbf{FNV} \\
\midrule
\textbf{0.50} & 0.0000 & -0.0001 & 0.0000 & 0.0000 & 0.0000 & 0.0000 \\
\textbf{0.75} & 0.0000 & 0.0002 & 0.0000 & 0.0000 & 0.0000 & 0.0000 \\
\textbf{1.00} & 0.0001 &  -0.0008 & 0.0000 & -0.0001 & -0.0007 & 0.0000 \\
\textbf{1.25} & 0.0002 & -0.0060 & 0.0000 &  -0.0003 & -0.0125 & 0.0000 \\
\textbf{1.50} & 0.0003 & -0.0233 & 0.0005 & -0.0009 & -0.0467 & 0.0001 \\
\textbf{1.75} & -0.0015 & -0.0594 & 0.0014 & -0.0016  & -0.1227 & 0.0080 \\
\bottomrule
\end{tabular}%
\vspace{0.5cm}
\caption{Absolute differences for the validation results of the FDG and PSMA models when all values in the predicted segmentation mask are set to zero where the PET values are below the specified threshold.}
\label{tab:suv}
\end{table*}

\subsection{Final submissions}
The results of Sec. \ref{sec:ana} have shown that a weighted multilabel approach has yielded the best performance in our cross-validation. 
Therefore, for the final model submission, we train two separate nnU-Net models using this approach on the FDG and PSMA datasets respectively. Given the demonstrated performance improvements with nnU-Net using residual encoders as reported in \cite{isensee2024nnu}, we adopt this architecture (ResEnc-M) in our work to leverage these benefits for our segmentation tasks. The FDG and PSMA models are trained for 1500 and 1000 epochs respectively. In contrast to the FDG model, longer training has not led to higher performance for the PSMA one. After manual examination of samples with high false positive volumes across different models from Tab. \ref{tab:ana_results}, we exclude one sample from the FDG and nine samples from the PSMA dataset. These samples either present certain special cases (e.g. \cite{grunig2021focal,Alam2016}) or may lack annotations. We assess that these samples are not essential for the model's performance.
The final results of this submission can be seen in  Tab. \ref{tab:final}. The FDG model has an increase of almost 2 \%, while the PSMA model has a slightly lower Dice score. Since we excluded multiple samples with high false-positive volumes, the FPV values are significantly lower in Tab. \ref{tab:final}. In comparison, the models using the weighted multilabel approach from Tab. \ref{tab:final} achieve higher FPV values of 8.15 for the FDG and PSMA datasets with excluded samples, respectively.
Since the PSMA model from Tab. \ref{tab:ana_results}  achieved the highest Dice score, we also submitted a variant that includes it and the FDG model from Tab. \ref{tab:final}.

\begin{table*}[t]
\centering
\small % Ensure the font size is applied here
\begin{tabular}{lccccccccc}
\toprule
\textbf{Method} & \multicolumn{3}{c}{\textbf{FDG Dataset}} & \multicolumn{3}{c}{\textbf{PSMA Dataset}} \\
\cmidrule(lr){2-4} \cmidrule(lr){5-7}
 & \textbf{DICE} & \textbf{FPV} & \textbf{FNV} & \textbf{DICE} & \textbf{FPV} & \textbf{FNV} \\
\midrule
\textbf{Multilabel} & 0.7690 & 3.85 & 6.95 & 0.6071 & 6.355 & 13.78 \\
\bottomrule
\end{tabular}%
\vspace{0.5cm}
\caption{Cross-validation results for the final submission.}
\label{tab:final}
\end{table*}